\documentclass[lettersize,journal]{IEEEtran}
\usepackage{amsmath,amsfonts}
\usepackage[ruled,linesnumbered]{algorithm2e}
\usepackage{amssymb}
\usepackage{array}
\usepackage[caption=false,font=normalsize,labelfont=sf,textfont=sf]{subfig}
\usepackage{textcomp}
\usepackage{stfloats}
\usepackage{url}
\usepackage{verbatim}
\usepackage{graphicx}
\usepackage{cite}
\usepackage{bm}
\usepackage{xcolor}     
\usepackage{booktabs}   
\usepackage{xspace}
\usepackage{orcidlink}
\usepackage{kotex}
\newcommand{\recifhe}{\mbox{Recifhe}\xspace}
\newcommand{\cheddar}{\mbox{Cheddar}\xspace}

\newcommand{\nbench}{12\xspace}
\newcommand{\noom}{3\xspace}
\newcommand{\nbenchpoly}{9\xspace}
\newcommand{\nslower}{3\xspace}
\newcommand{\polyspvsctxt}{0.92$\times$\xspace}
\newcommand{\spvsctxt}{1.25$\times$\xspace}
\newcommand{\spvspoly}{1.36$\times$\xspace}

\newcommand{\polymemvsctxt}{1.23$\times$\xspace}
\newcommand{\memvsctxt}{1.14$\times$\xspace}
\newcommand{\memvspoly}{0.93$\times$\xspace}

\newcommand{\rescale}{\mbox{\texttt{Rescale}}\xspace}
\newcommand{\bootstrap}{\mbox{\texttt{Bootstrap}}\xspace}

\newcommand{\addcc}{\mbox{\texttt{AddCC}}\xspace}
\newcommand{\addcp}{\mbox{\texttt{AddCP}}\xspace}
\newcommand{\mulcc}{\mbox{\texttt{MulCC}}\xspace}
\newcommand{\mulcp}{\mbox{\texttt{MulCP}}\xspace}
\newcommand{\rotate}{\mbox{\texttt{Rotate}}\xspace}

\newcommand{\modup}{\texttt{ModUp}\xspace}
\newcommand{\moddown}{\texttt{ModDown}\xspace}

\newcommand{\automorphism}{\texttt{Automorphism}\xspace}
\newcommand{\keyip}{\texttt{KIP}\xspace}
\newcommand{\invmoddown}{\texttt{invModDown}\xspace}
\newcommand{\pmulcp}{\texttt{pMulCP}\xspace}
\newcommand{\pmulcc}{\texttt{pMulCC}\xspace}
\newcommand{\padd}{\texttt{pAdd}\xspace}

\begin{document}

\title{Ciphertext- and Polynomial-Level Optimization \\ for Fully Homomorphic Encryption}

\author{
    Seongho Kim\,\orcidlink{0009-0008-9306-9301}, 
    Heelim Choi\,\orcidlink{0000-0002-1885-0578}, 
    Jaemin Kim\,\orcidlink{0009-0005-6444-991X}, 
    Seonyoung Cheon\,\orcidlink{0009-0005-3463-716X}, 
    Dongkwan Kim\,\orcidlink{0000-0002-9611-0233}, 
    Jaeho Lee\,\orcidlink{0000-0002-2735-0647}, 
    Hoyun Youm\,\orcidlink{0009-0005-6623-3995}, 
    Dongyoon Lee\,\orcidlink{0000-0002-2240-3316} \IEEEmembership{Member, IEEE}, 
    Hanjun Kim\,\orcidlink{0000-0002-0762-7901} \IEEEmembership{Member, IEEE}, 
    Yongwoo Lee\,\orcidlink{0000-0002-7458-8885} \IEEEmembership{Member, IEEE}
}





\maketitle
\begin{abstract}
Fully homomorphic encryption (FHE) schemes such as RNS-CKKS enable privacy-preserving services through direct computation on encrypted data. While recent FHE compilers optimize FHE programs, they operate at the coarse-grained ciphertext level, where each ciphertext operation comprises a sequence of polynomial operations. At this granularity, the compilers miss polynomial-level optimization opportunities across ciphertext operations. This work presents Recifhe, a new multi-level compiler that supports both ciphertext-level and polynomial-level optimization. At the ciphertext level, Recifhe transforms a non-FHE input program into an FHE program by inserting ciphertext management operations and applies global optimizations. At the polynomial level, Recifhe eliminates redundant polynomial computations across ciphertext operations. Recifhe achieves a 1.25$\times$ speedup over ciphertext-level-only optimization.

\end{abstract}

\begin{IEEEkeywords}
FHE compiler, RNS-CKKS, ciphertext-level optimization, polynomial-level optimization
\end{IEEEkeywords}

\section{Introduction}
\label{sec:introduction}
Fully homomorphic encryption (FHE) enables direct computation on encrypted data without decryption, supporting privacy-preserving services.
Among FHE schemes, RNS-CKKS~\cite{rnsckks:2018:cheon} is suited to machine learning because it supports fixed-point arithmetic on real numbers and SIMD vectorization.
To utilize RNS-CKKS, programmers must transform a non-FHE program into an FHE program by inserting ciphertext management operations that adjust ciphertext attributes.

To reduce the programming burden, recently proposed FHE compilers~\cite{hecate:2022:lee, elasm:2023:lee, reserve:2024:lee, eva:2020:dathathri, dacapo:2024:cheon, halo:2025:cheon, orion:2025:ebel, HEaaN.MLIR:2023:park, antace:2025:Li} automatically place ciphertext management operations and optimize FHE programs, but they fail to fully exploit optimization opportunities.
Most compilers~\cite{hecate:2022:lee, elasm:2023:lee, reserve:2024:lee, eva:2020:dathathri, dacapo:2024:cheon, halo:2025:cheon, orion:2025:ebel} operate at the ciphertext level, where each ciphertext operation comprises a sequence of polynomial operations, thereby missing fine-grained optimization opportunities such as eliminating redundant polynomial operations across ciphertext operations.
A few compilers~\cite{HEaaN.MLIR:2023:park, antace:2025:Li} lower ciphertext operations into polynomial sequences, but their optimization is limited to a single ciphertext operation because they do not track the polynomial attributes across ciphertext operations.

This work introduces \recifhe, a new multi-level compiler that supports both ciphertext-level and polynomial-level optimizations, each suited to its granularity.
At the ciphertext level, \recifhe transforms a non-FHE input program into an FHE program and applies existing global optimizations such as \rescale hoisting.
At the polynomial level, \recifhe eliminates redundant polynomial computations across ciphertext operations through common subexpression elimination, performance-aware \moddown hoisting, and operation fusion.
The profitability of \moddown hoisting depends on the dataflow, so \recifhe profiles the latency of each polynomial operation and applies hoisting only when the estimated latency benefit exceeds the overhead.
\recifhe also reduces the memory growth inherent in polynomial-level optimization, caused by the extended polynomial lifetimes, through liveness-driven operation scheduling.
As a result, \recifhe achieves a \spvsctxt speedup over ciphertext-level-only optimization, and a \spvspoly speedup with \memvspoly the memory footprint over the manually optimized polynomial sequences in an existing FHE library.

\section{Background and Motivation}
\label{sec:bg}

The residue number system variant of CKKS (RNS-CKKS) \cite{rnsckks:2018:cheon} encodes a vector of real numbers into a plaintext polynomial in $\mathbb{Z}[X]/(X^N+1)$, where $N$ is the polynomial modulus degree, providing $N/2$ fixed-point slots with SIMD semantics.
RNS-CKKS encrypts the plaintext as a pair of ciphertext polynomials in $\mathbb{Z}_Q[X]/(X^N+1)$, where the coefficient modulus $Q$ bounds the coefficients.
Each polynomial consists of up to $L$ residue polynomials with moduli $q_i$, where $Q = \prod^L_{i=1} q_i$.
The GPU library~\cite{cheddar:2024:choi} used in this work adopts $N = 2^{16}$ and $L = 27$ with 32-bit moduli, so a plaintext occupies $4\,\text{bytes} \times N \times L = 6.75$\,MB, and a ciphertext twice as much.


Recent FHE compilers~\cite{hecate:2022:lee, reserve:2024:lee, eva:2020:dathathri, elasm:2023:lee, dacapo:2024:cheon, orion:2025:ebel, halo:2025:cheon} operate at the ciphertext level, managing two attributes of a ciphertext, the \emph{level} and the \emph{scale}.
The level is the number of \rescale operations applicable to the ciphertext, up to $L$, and the scale is the fixed-point scaling factor by which the encoded message value is multiplied.
A multiplication on a ciphertext enlarges the scale of the result, potentially incurring scale overflow.
\rescale divides the scale by the last modulus and lowers the level by one, preventing the overflow.
\bootstrap restores the depleted level to its maximum of $L$.

The ciphertext-level compilers automatically place the ciphertext management operations, \rescale and \bootstrap.
The scale management compilers~\cite{hecate:2022:lee, reserve:2024:lee, eva:2020:dathathri, elasm:2023:lee} insert \rescale operations to prevent scale overflow.
The bootstrapping management compilers~\cite{dacapo:2024:cheon, orion:2025:ebel, halo:2025:cheon} insert \bootstrap operations that recover the level, supporting programs with long multiplication chains while minimizing the \bootstrap count.
These compilers also apply global optimizations, such as \rescale hoisting~\cite{reserve:2024:lee, orion:2025:ebel}, to reduce the ciphertext operation count but fail to remove redundant polynomial computations across ciphertext operations.

\begin{figure*}[t]
\centering
    \includegraphics[width=\hsize]{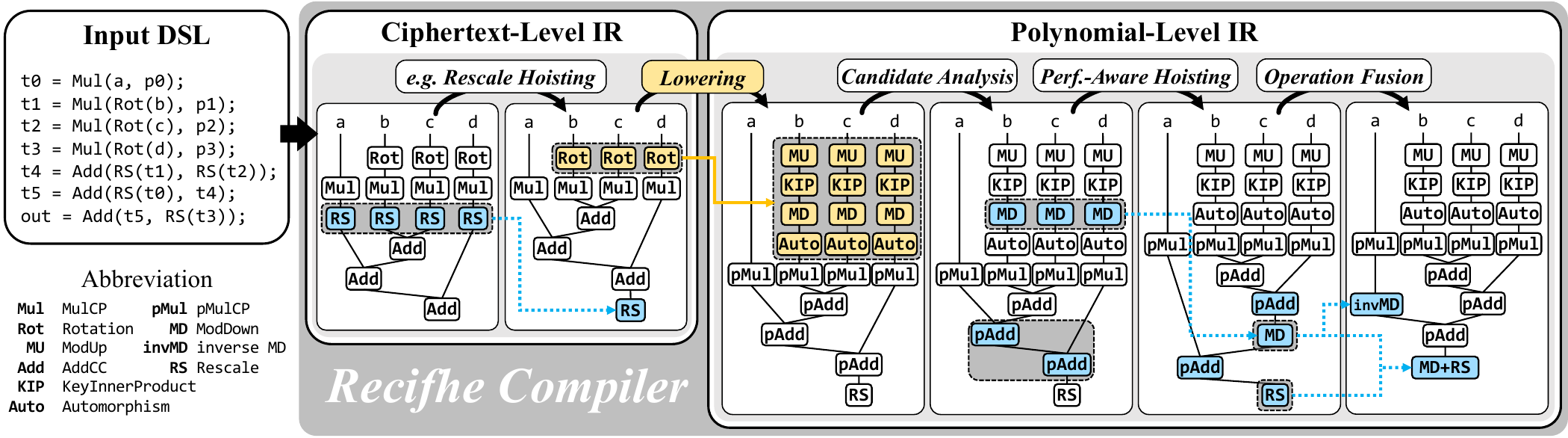}
    \caption{Overview of \recifhe with the running example. The plaintexts (p0--p3) of the input DSL are omitted for simplicity.}
    \label{fig:overview:compiler}
\end{figure*}


Each ciphertext operation comprises a sequence of polynomial operations, and each polynomial carries a third attribute, the \emph{RNS basis}.
The RNS basis is the set of moduli over which a polynomial is represented, and a polynomial operation requires its operands to share the same RNS basis.
Table~\ref{tab:bg:map} lists the polynomial operation sequence of each ciphertext operation, omitting \rescale and \bootstrap, which appear identically at both levels.
\mulcc and \rotate produce an intermediate result under a different key, so key-switching restores it to the original key via \modup, \keyip, and \moddown.
\keyip computes the inner product of the \modup result with the key-switching key, amplifying the key-switching error, so RNS-CKKS temporarily extends the working modulus from $Q$ to $PQ$ with an auxiliary modulus $P = \prod_{j} p_j$.
\modup lifts a polynomial from the $Q$ basis to the $PQ$ basis by adding the $p_j$ moduli to its basis, and after \keyip, \moddown divides the result by $P$ to reduce the key-switching noise and returns it to the $Q$ basis.
\automorphism rotates the slots through an index permutation.
The \emph{RNS-basis-insensitive} operations in Table~\ref{tab:bg:sensitivity} apply identically on both the $Q$ and $PQ$ bases, processing each residue polynomial independently and preserving the factor $P$ introduced by key switching.
\modup, \moddown, \rescale, and \bootstrap, which change the RNS basis, internally use the number theoretic transform (NTT) and its inverse (iNTT) to perform basis conversion.
Fusing an adjacent pair of these operations can cancel out the internal transformations.

\begin{table}[t]
    \centering
    \caption{Ciphertext-to-Polynomial Operation Mapping}
    \label{tab:bg:map}
    \small
    \begin{tabular}{@{}c p{0.7\columnwidth}@{}}
    \toprule
    Ciphertext Op. & Polynomial Op. \\
    \midrule
    \addcc, \addcp & \texttt{pAdd} \\
    \mulcp & \texttt{pMulCP} \\
    \mulcc & \texttt{pMulCC}$\mathbin{\rightarrow}$\texttt{ModUp}$\mathbin{\rightarrow}$\texttt{KIP}$\mathbin{\rightarrow}$\texttt{ModDown} \\
    \rotate & \texttt{ModUp}$\mathbin{\rightarrow}$\texttt{KIP}$\mathbin{\rightarrow}$\texttt{ModDown}$\mathbin{\rightarrow}$\texttt{Automorphism} \\
    \bottomrule
    \end{tabular}
\end{table}

A few compilers~\cite{HEaaN.MLIR:2023:park, antace:2025:Li} decompose ciphertext operations into polynomial operation sequences but do not manage the RNS basis across ciphertext operations, limiting their optimization to a single ciphertext operation.
An existing FHE library~\cite{cheddar:2024:choi} manually implements optimizations across ciphertext operations, such as removing repeated \moddown operations.
The library optimization applies a fixed algorithm tuned to a specific dataflow, so it neither adapts to new programs nor accounts for the trade-off between the removed \moddown operations and the operations raised to the $PQ$ basis.
Furthermore, the library optimization extends polynomial lifetimes, incurring high memory overhead.
These limitations motivate a compiler that applies dataflow-adaptive optimization at the polynomial level.


\begin{table}[t]
    \centering
    \caption{RNS-Basis Sensitivity of Polynomial Operations}
    \label{tab:bg:sensitivity}
    \small
    \begin{tabular}{@{}cp{0.7\columnwidth}@{}}
    \toprule
    Insensitive & \padd, \pmulcp, \automorphism \\
    Sensitive & \modup, \keyip, \moddown, \newline \pmulcc, \rescale, \bootstrap \\
    \bottomrule
    \end{tabular}
\end{table}


\section{\recifhe Compiler}
\label{sec:implementation}

\recifhe applies each optimization at the granularity suited to it, adopting a multi-level design that performs global optimization on a
ciphertext-level intermediate representation (IR) and fine-grained optimization across ciphertext operations on a polynomial-level IR.
On the ciphertext-level IR, as shown in Fig.~\ref{fig:overview:compiler}, \recifhe transforms an input Python DSL program into an FHE program and applies existing global optimizations, including the ciphertext management operation placement and \rescale hoisting.
Applying global optimizations at the coarse-grained ciphertext level allows \recifhe to reduce optimization complexity compared to optimizing the entire program directly at the fine-grained polynomial level.
\recifhe then decomposes each ciphertext operation into its corresponding polynomial sequence, enabling optimizations across ciphertext operations.
\recifhe identifies candidates for \moddown hoisting on the polynomial-level IR through dataflow analysis, applies performance-aware hoisting guided by offline profiling, fuses operations, and schedules polynomial operations to reduce memory growth.

Fig.~\ref{fig:overview:compiler} illustrates \recifhe on a running example that computes $ y = p_0 \cdot a + p_1 \cdot \mathrm{rot}(b, r_1) + p_2 \cdot \mathrm{rot}(c, r_2) + p_3 \cdot \mathrm{rot}(d, r_3)$, where each rotated ciphertext is multiplied by a plaintext and the resulting products are accumulated into a single output.
This rotate-multiply-accumulate pattern recurs in FHE matrix multiplication algorithms~\cite{double:2020:bossuat, orion:2025:ebel}, where each rotation lowers into \modup, \keyip, \moddown, and \automorphism.
The results of the three \moddown operations flow through only RNS-basis-insensitive operations, so hoisting them into a single \moddown reduces the \moddown count and enables fusing the surrounding operations.

\begin{algorithm}[t]
\small
\DontPrintSemicolon
\SetKwProg{Fn}{Function}{}{end}
\SetKw{Continue}{continue}
\KwIn{Polynomial-level function DAG $\Pi$, Latency profile $\mathcal{E}$}
\KwOut{\moddown{}-hoisted DAG $\Pi$}
\Fn{\textnormal{PerformanceAwareModDownHoisting} ($\Pi$, $\mathcal{E}$) :}
{
     $S = \{op \in \Pi \mid \textnormal{isSensitive}(op) \wedge op \neq \moddown \}$ \; 
     $I = \{op \in \Pi \mid \textnormal{isInsensitive}(op) \vee op = \moddown \}$ \;
    \ForEach{$s \in S$}
    {
        $B \gets \{ b \in S \mid \exists \pi \in \textnormal{Path}(b,s) : \textnormal{ops}(\pi) \subseteq I \}$\label{alg:hoist:analysis:start}\label{alg:hoist:analysis:boundary}\;
        $R \gets \bigcup_{b \in B} \bigcup_{\pi \in \textnormal{Path}(b,s)} \textnormal{ops}(\pi)$\label{alg:hoist:analysis:region}\;
        $M \gets \{ op \in R \mid op = \moddown \}$\label{alg:hoist:analysis:target}\;
        \If{$|M| < 2$}
        {
            \Continue
        }\label{alg:hoist:analysis:end}
        $F \gets \textnormal{findNearestSensitiveSuccessors}(\Pi, M)$\label{alg:hoist:hoist:start}\label{alg:hoist:hoist:frontier}\;
        $L \gets \{op \in \textnormal{Path}(m,f) \mid m \in M, f \in F\}$\label{alg:hoist:hoist:raise}\;
        $\textit{benefit} \gets \sum_{m \in M} \mathcal{E}(m)
            - \sum_{f \in F} \mathcal{E}(\moddown_{f})$\label{alg:hoist:hoist:benefit}\;
        $\textit{overhead} \gets \sum_{op \in L}
            \big(\mathcal{E}_{PQ}(op) - \mathcal{E}_{Q}(op)\big)$\label{alg:hoist:hoist:overhead}\;
        \If{$\textit{benefit} > \textit{overhead}$\label{alg:hoist:hoist:cond}}
        {
            $\Pi \gets \textnormal{reassociate}(\Pi, R, M)$\label{alg:hoist:hoist:reassoc}\;
            $\Pi \gets \textnormal{removeModDown}(\Pi, M)$\label{alg:hoist:hoist:remove}\;
            $\Pi \gets \textnormal{insertModDownBefore}(\Pi, F)$\label{alg:hoist:hoist:add}\;
        }\label{alg:hoist:hoist:end}
    }
    \KwRet{$\Pi$}\;
}
\caption{Performance-Aware \moddown{} Hoisting.}
\label{alg:hoist}
\end{algorithm}


\textbf{Rescale Hoisting and Lowering.}
\recifhe first places \bootstrap operations and manages the scale of the program using existing ciphertext-level techniques~\cite{hecate:2022:lee, reserve:2024:lee, eva:2020:dathathri, elasm:2023:lee, dacapo:2024:cheon, halo:2025:cheon, orion:2025:ebel}, so the level and the scale of every value are known by the time \recifhe lowers the ciphertext operations.
\recifhe inherits \rescale hoisting~\cite{reserve:2024:lee, orion:2025:ebel} and maintains this global transformation at the ciphertext level, since the program contains more operations at the polynomial level.
\recifhe hoists \rescale operations past any operation that preserves the level and the scale, merging them to reduce the \rescale count.
\recifhe then lowers each ciphertext operation into its polynomial operation sequence.
The lowering translates each ciphertext operation independently, so \recifhe also eliminates repeated polynomial operations for the same input polynomial across the lowered sequences through common subexpression elimination.

\textbf{Candidate Analysis.}
\recifhe analyzes the polynomial-level DAG of each function to find optimization candidates where multiple \moddown operations can be hoisted and merged into fewer operations (Algorithm~\ref{alg:hoist} Lines \ref{alg:hoist:analysis:start}-\ref{alg:hoist:analysis:end}).
\recifhe visits each RNS-basis-sensitive operation $s$ other than \moddown, leveraging that the earlier ciphertext-level optimizations already place the ciphertext operations that lower into the sensitive operations.
To identify the optimizable region, the analysis first finds the boundary operation set $B$ (Line \ref{alg:hoist:analysis:boundary}), which consists of the preceding sensitive operations other than \moddown that can reach $s$ only through RNS-basis-insensitive and \moddown operations. 
The optimizable region $R$ consists of the intermediate operations on every path from a boundary operation to $s$ (Line \ref{alg:hoist:analysis:region}).
Finally, the analysis builds the hoisting target set $M$, the \moddown operations in the region (Line \ref{alg:hoist:analysis:target}). 
Their results can flow toward $s$ only through insensitive operations.
The insensitive operations compute correctly on both the $Q$ and $PQ$ bases, so a hoisted \moddown crosses them without changing the encoded values.
\recifhe takes the region as a candidate when $M$ holds multiple \moddown operations.
The regions of different sensitive operations may be nested, complicating the analysis. 
\recifhe handles the nested regions with iterative optimization, because hoisting in a nested region and its enclosing region can be composed without conflict.


\textbf{Performance-Aware Hoisting.}
Hoisting removes the \moddown operations in $M$ of each candidate region and places one hoisted \moddown before each operation in the sensitive frontier $F$ (Algorithm~\ref{alg:hoist} Lines~\ref{alg:hoist:hoist:start}-\ref{alg:hoist:hoist:end}).
Function \textnormal{findNearestSensitiveSuccessors} finds $F$ (Line~\ref{alg:hoist:hoist:frontier}), the set of sensitive operations that the results of $M$ first reach via RNS-basis-insensitive operations only, and $\moddown_f$ denotes the hoisted \moddown before each $f \in F$.
The set $L$ consists of the insensitive operations on the paths from $M$ to $F$, which process $PQ$-basis polynomials after hoisting (Line~\ref{alg:hoist:hoist:raise}).
The $PQ$ basis holds more moduli than the $Q$ basis, so the operations in $L$ incur more computation.
Whether this added cost outweighs the savings from the removed \moddown operations depends on the dataflow of each region.
\recifhe therefore profiles the latency $\mathcal{E}$ of every polynomial operation offline at each level and RNS basis, with the basis denoted as a subscript, and estimates the cost of a set of operations as the sum of their latencies.
\recifhe accepts a hoist only when the benefit, the latency of $M$ minus the latency of the hoisted \moddown operations at $F$, exceeds the overhead, the latency increase of $L$ raised to the $PQ$ basis (Lines~\ref{alg:hoist:hoist:benefit}-\ref{alg:hoist:hoist:cond}).
\recifhe applies an accepted hoist by reordering the accumulations in the region with function \textnormal{reassociate} to gather the results of $M$, removing $M$, and inserting $\moddown_f$ before each $f \in F$ (Lines~\ref{alg:hoist:hoist:reassoc}-\ref{alg:hoist:hoist:add}).
The reordering preserves the encoded result because modular addition is associative and commutative.

\textbf{Operation Fusion.}
\recifhe fuses adjacent RNS-basis-sensitive operations, such as a \moddown and a \rescale, to remove an iNTT and NTT pair if there is no operation between them.
The fusible pair may not be adjacent, as in the running example where an addition separates a \moddown from a \rescale.
To handle this case, \recifhe raises the other operand of the addition, which remains on the $Q$ basis, to the $PQ$ basis by inserting an \invmoddown, allowing the \moddown to sink past the addition and become adjacent to the \rescale.
\invmoddown is a lightweight lift that multiplies each residue polynomial in the $Q$ basis by $P$, satisfying $\moddown(\invmoddown(a))=a$.

\textbf{Polynomial Scheduling.}
Optimization at the polynomial level inherently increases memory pressure, and \recifhe reduces the memory growth with polynomial scheduling.
The lowering places the polynomial operations of each ciphertext operation consecutively, separating a value's producer from its consumers, and hoisting and fusion further extend polynomial lifetimes across ciphertext boundaries.
\recifhe therefore moves each operation to immediately follow its most recently defined operand, shrinking the operand's live range when the operation is its last use.
\recifhe then reuses the buffer of a dead value for a later result of the same size, reducing the number of polynomials that are live simultaneously.



\begin{figure*}[t]
    \captionsetup[subfloat]{labelfont=footnotesize,textfont=footnotesize}
    \centering
    \subfloat[Latency comparison. Lower is better.]{
    \includegraphics[width=0.48\linewidth]{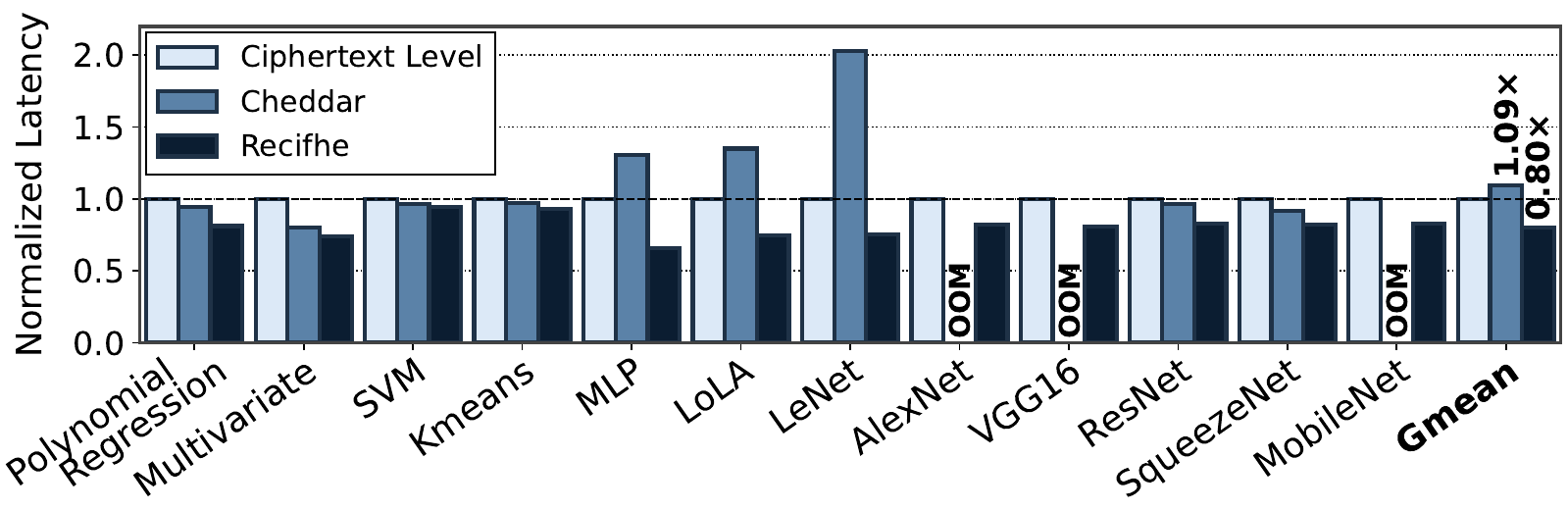}
    \label{fig:evaluation:speedup}
    }
    \hfill
    \subfloat[Memory footprint comparison. Lower is better.]{
    \includegraphics[width=0.48\linewidth]{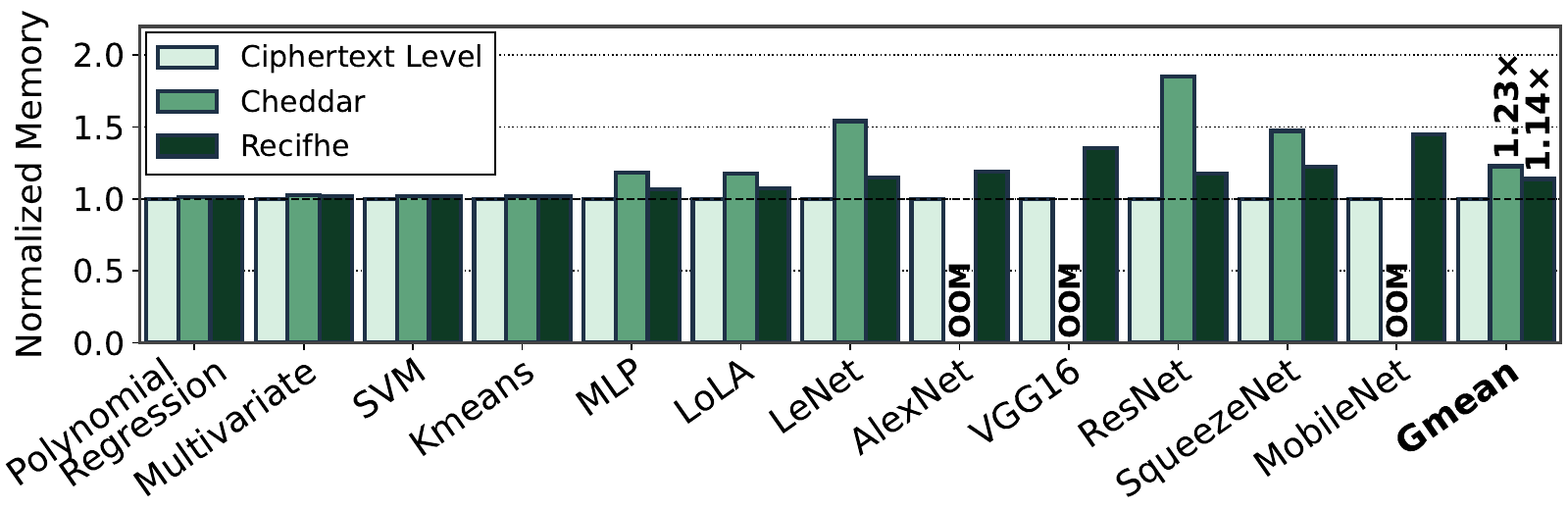}
    \label{fig:evaluation:memory}
    }
    \caption{Comparison of the Cheddar~\cite{cheddar:2024:choi} baseline and \recifhe over the ciphertext-level-only baseline.}
    \label{fig:evaluation}
\end{figure*}


\section{Evaluation}
\label{sec:eval}
\textbf{Environment.}
\recifhe is built on the MLIR-based Hecate framework~\cite{hecate:2022:lee} with a Python DSL frontend, using \cheddar~\cite{cheddar:2024:choi} as the GPU backend target.
\recifhe is retargetable to other backend libraries by re-profiling operation latencies.
\recifhe is evaluated on \nbench benchmarks against two baselines sharing the same frontend and backend.
The ciphertext-level-only baseline in Fig.~\ref{fig:evaluation} applies prior ciphertext-level optimizations~\cite{dacapo:2024:cheon, reserve:2024:lee, orion:2025:ebel, hecate:2022:lee, elasm:2023:lee, halo:2025:cheon, eva:2020:dathathri}, and the \cheddar baseline adds the manually optimized polynomial sequences on top of that baseline.
All three configurations share the same RNS-CKKS setting, with the polynomial modulus degree $N=2^{16}$, the default scale of $2^{40}$, and the \cheddar \bootstrap modulus configuration at the 128-bit security level, yielding an effective multiplicative depth of $13$ between \bootstrap operations.
All benchmarks use pre-encoded plaintexts and place \bootstrap operations following prior benchmark suites~\cite{halo:2025:cheon, dacapo:2024:cheon, orion:2025:ebel}.
The experiments run on two Intel Xeon Gold 6326 CPUs with 512\,GB RAM and an NVIDIA RTX PRO 6000 Blackwell Max-Q GPU with 96\,GB VRAM.

\textbf{Analysis.}
Fig.~\ref{fig:evaluation:speedup} reports the latency normalized to the ciphertext-level-only baseline, and Fig.~\ref{fig:evaluation:memory} reports the memory footprint.
The \cheddar baseline accelerates 6 of the \nbenchpoly benchmarks it completes but slows down on the other \nslower, because the library optimization raises operations into the heavier $PQ$ basis regardless of whether the dataflow makes the hoist profitable, exactly the fixed-algorithm limitation described in \S\ref{sec:bg}.
The slowdown on the \nslower benchmarks lowers the average to \polyspvsctxt the speed of the ciphertext-level-only baseline.
The same unconditional transformation also drives the \cheddar baseline's memory cost, since it lacks a scheduling pass to reclaim the extended live ranges.
Its $PQ$-basis polynomials thus remain live longer than necessary, using more memory than the ciphertext-level-only baseline on every benchmark it completes, \polymemvsctxt the footprint on average, and running out of memory on \noom of the \nbench benchmarks.
The ResNet result shows that \recifhe faithfully reimplements the library optimization~\cite{cheddar:2024:choi}, reproducing \cheddar's artifact result within 0.9\% by running in 0.714\,s compared to the artifact's 0.720\,s.

In contrast, \recifhe applies hoisting only when Algorithm~\ref{alg:hoist} finds it profitable, reaching \spvsctxt speedup over the ciphertext-level-only baseline across all \nbench benchmarks and \spvspoly over the \cheddar baseline on the \nbenchpoly benchmarks the \cheddar baseline completes.
The liveness-driven scheduling of \recifhe reduces the memory consumption from hoisting and fusion, holding the footprint to \memvsctxt that of the ciphertext-level-only baseline and \memvspoly that of the \cheddar baseline.
These gains cost additional compile time, with the polynomial-level passes taking 39.0\% of \recifhe's compile time on average, and performance-aware hoisting and fusion taking 1.31\% and 14.3\% of the total compile time, respectively.
Compile times range from 0.0182\,s (Polynomial Regression) to 22.4\,s (MobileNet), averaging 5.14\,s across all \nbench benchmarks, 2.92$\times$ that of the ciphertext-level-only baseline in geometric mean.
\section{Conclusion}
\label{sec:conclusion}

This work proposes \recifhe, a new multi-level compiler that automatically generates FHE programs with global optimizations at the ciphertext level and removes the redundant polynomial computations across ciphertext operations.
Guided by measured per-operation latencies, \recifhe applies \moddown hoisting only when the estimated benefit exceeds the overhead, and reduces the resulting memory growth through liveness-driven scheduling.
As a result, \recifhe achieves a \spvsctxt speedup over ciphertext-level-only optimization.

\bibliographystyle{IEEEtran}
\bibliography{ref}

@inproceedings{hecate:2022:lee,
    author = {Lee, Yongwoo and others},
    title = {{HECATE}: {Performance-Aware Scale Optimization for Homomorphic Encryption Compiler}},
    year = {2022},
    booktitle = {Proc. IEEE/ACM Int. Symp. Code Gener. Optim.},  
}

@inproceedings{elasm:2023:lee,
    title={{ELASM}: Error-Latency-Aware Scale Management for Fully Homomorphic Encryption},
    author={Lee, Yongwoo and others},
    booktitle={Proc. USENIX Secur. Symp.},
    year={2023},
}

@inproceedings{dacapo:2024:cheon,
    author = {Seonyoung Cheon and others},
    title = {{DaCapo}: Automatic Bootstrapping Management for Efficient Fully Homomorphic Encryption},
    year = {2024},
    booktitle = {Proc. USENIX Secur. Symp.},
}

@inproceedings{reserve:2024:lee,
    title = {Performance-aware scale analysis with reserve for homomorphic encryption},
    author={Yongwoo Lee and others},
    booktitle = {Proc. ACM Int. Conf. on Archit. Support for Prog. Lang. Oper. Syst., vol. 1},
    year = {2024},
}

@inproceedings{halo:2025:cheon,
    author = {Cheon, Seonyoung and others},
    year = {2025},
    booktitle = {Proc. ACM Int. Conf. on Archit. Support for Prog. Lang. Oper. Syst., vol. 1},
    title = {{HALO}: Loop-aware Bootstrapping Management for Fully Homomorphic Encryption},
}

@inproceedings{rnsckks:2018:cheon,
    author={Cheon, Jung Hee and others},
    title={A Full RNS Variant of Approximate Homomorphic Encryption},
    booktitle={Proc. Sel. Areas in Cryptog.},
    year={2018},
}

@article{HEaaN.MLIR:2023:park,
    author = {Park, Sunjae and others},
    title = {HEaaN.MLIR: An Optimizing Compiler for Fast Ring-Based Homomorphic Encryption},
    year = {2023},
    journal = {Proc. ACM Program. Lang.},
}

@inproceedings{eva:2020:dathathri,
    title = {{EVA}: An Encrypted Vector Arithmetic Language and Compiler for Efficient Homomorphic Computation},
    booktitle = {Proc. ACM SIGPLAN Conf. on Program. Lang. Design Impl.},
    author = {Dathathri, Roshan and others},
    year = {2020},
}

@inproceedings{orion:2025:ebel,
    author={Austin Ebel and Karthik Garimella and Brandon Reagen},
    year = {2025},
    booktitle = {Proc. ACM Int. Conf. on Archit. Support for Prog. Lang. Oper. Syst., vol. 1},
    title={Orion: A Fully Homomorphic Encryption Framework for Deep Learning},
}

@inproceedings{antace:2025:Li,
    author = {Li, Long and others},
    title = {ANT-ACE: An FHE Compiler Framework for Automating Neural Network Inference},
    year = {2025},
    booktitle = {Proc. ACM/IEEE Int. Symp. Code Gener. and Optim.},
}

@inproceedings{cheddar:2024:choi,
  title = {Cheddar: A Swift Fully Homomorphic Encryption Library Designed for GPU Architectures},
  author = {Choi, Wonseok and Kim, Jongmin and Ahn, Jung Ho},
  year = {2025},
  booktitle = {Proc. ACM Int. Conf. on Archit. Support for Prog. Lang. Oper. Syst., vol. 1},
}

@InProceedings{double:2020:bossuat,
    author="Bossuat, Jean-Philippe and others",
    title="Efficient Bootstrapping for Approximate Homomorphic Encryption with Non-sparse Keys",
    booktitle="Proc. Adv. Cryptol. -- EUROCRYPT, Part I",
    year="2021",
}

\end{document}